**Ukrainian wartime astronomy and its prospects**

*The Russian invasion of Ukraine damaged or compromised astronomical facilities and has prompted the displacement of researchers. A plan to restore Ukrainian astronomy, rooted in a deeper integration with the international community, is now being developed.*

Danilo Albergaria, Leiden University, albergaria@strw.leidenuniv.nl – corresponding author (Leiden, the Netherlands)

Kateryna Frantseva, Leiden University, frantseva@strw.leidenuniv.nl (Leiden, the Netherlands)

Pedro Russo, Leiden University (Leiden, the Netherlands)

Svitlana Babiichuk, National Center "Junior Academy of Sciences of Ukraine" (Kyiv, Ukraine)

Oksana Berezhna, Ministry of Education and Science of Ukraine (Kyiv, Ukraine)

Sofiia Denyshchenko, V.N. Karazin Kharkiv National University (Kharkiv, Ukraine)

Daria Dobrycheva, Main Astronomical Observatory, NAS of Ukraine (Kyiv, Ukraine)

Vadym Kaydash, Institute of Astronomy, V.N. Karazin Kharkiv National University (Kharkiv, Ukraine)

Olena Kompaniiets, Main Astronomical Observatory, NAS of Ukraine (Kyiv, Ukraine)

Oleksander Konovalenko, Institute of Radio Astronomy, NAS of Ukraine (Kharkiv, Ukraine)

Yurii Kulinich, Astronomical Observatory, Ivan Franko National University of Lviv (Lviv, Ukraine)

Igor Lukyanyk, Astronomical Observatory, Taras Shevchenko National University of Kyiv (Kyiv, Ukraine)

Vladyslava Marsakova, a) Richelieu Science lyceum, Odesa b) Main Astronomical Observatory, NAS of Ukraine c) Odesa National Marine University (Odesa, Ukraine)

Bohdan Novosyadlyj, Astronomical Observatory, Ivan Franko National University of Lviv (Lviv, Ukraine)

Elena Panko, Odesa I.I. Mechnikov National University (Odesa, Ukraine)

Volodymyr Reshetnyk, Taras Shevchenko National University of Kyiv (Kyiv, Ukraine)

Ivan Slyusarev, Department of Astronomy and Space Informatics, V.N. Karazin Kharkiv National University (Kharkiv, Ukraine)

Iurii Sushch, a) Astronomical Observatory, Ivan Franko National University of Lviv b) Centro de Investigaciones Energéticas, Medioambientales y Tecnológicas (CIEMAT), E-28040 Madrid, Spain (Lviv, Ukraine)

Ganna Tolstanova, Taras Shevchenko National University of Kyiv (Kyiv, Ukraine)

Iryna Vavilova, Main Astronomical Observatory, NAS of Ukraine (Kyiv, Ukraine)

Liubov Yankiv-Vitkovska, Lviv Polytechnic National University (Lviv, Ukraine)

Yaroslav Yatskiv, Main Astronomical Observatory, NAS of Ukraine (Kyiv, Ukraine)

Vyacheslav Zakharenko, Institute of Radio Astronomy, NAS of Ukraine, Kharkiv (Kharkiv, Ukraine)

Science is not immune to the devastation of war. Since the beginning of Russia's full-scale invasion in February 2022, Ukrainian science has faced severe setbacks. Educational and research institutions have been [systematically targeted](#) and heavily damaged, particularly in regions near the frontlines or under occupation. Disruptions of research and education are direct consequences of aggressions that have deliberately undermined Ukraine's civilian infrastructure and intellectual life. Although the full extent of the destruction remains unclear, some details have emerged.

A March 2024 study by UNESCO and the Junior Academy of Sciences of Ukraine reported 1,443 damaged buildings at 177 institutions, with US$1.26 billion needed—mostly for university reconstruction[1]. Between the start of the war and early 2024, academic salaries had dropped 39% and Ukraine's public R&D budget was halved[1]. By April 2024, 12% of the 88,629 researchers and university professors were forced to emigrate or were internally displaced[1]. A December 2023 study showed 23.5% of remaining scientists lost key research resources, and 20.7% lacked physical access to their institutions—either due to relocation or online-only access[2]. By the end of 2022, Ukraine had already lost an estimated 20% of its research capacity[2].

Before the Russian invasion, Ukrainian scientific institutions enjoyed substantial international recognition. Alongside space research and technology, astronomy is a field in which Ukraine has historically excelled. Ukrainian astronomers' notable contributions in the last two decades include the discovery and characterization of exoplanets and Solar System bodies, massive star formation, variable stars, pulsars, abundance of carbon in the interstellar medium using radio spectroscopy, stellar and galactic chemical evolution, cosmological modelling in four or multidimensional space-time, large-scale structure of the Universe, and machine learning applications[3, 4, 5, 6, 7, 8]. Ukraine also has long-standing expertise in space surveillance and tracking, supported by a wide telescope network for astrometric and photometric observations.

More recently, Ukrainians were responsible for the detection of several exocomets around Beta Pictoris[9] and the development of a new machine learning approach to uncover cometary bodies around other stars in the dataset of the Transiting Exoplanet Survey Satellite (TESS), which identified 20 potential candidates[10]. Ukrainian scientists have continued publishing on topics ranging from stellar kinematics and exoplanetary systems to globular clusters, asteroids, and lunar evolution.

Once vibrant, Ukraine's astronomy community suffered significant losses: damaged infrastructure, destroyed or looted equipment, disrupted research projects, and lost human resources. During the meeting *[Recovery Plan for Ukrainian Astronomy](#)*, held on 10th and 11th June at the Leiden Observatory in the Netherlands, leading Ukrainian astronomers described the impacts of the war to their Dutch and European counterparts, setting the stage for upcoming rebuilding plans. We navigate their losses in the next two sections and briefly comment on the event's importance for the future of Ukrainian astronomy in the closing segment.

**Research Facilities Losses**

Losses for Ukrainian astronomy date back to 2014, when Russia annexed Crimea. The peninsula is home to some key observational facilities, like the Shine Mirror Telescope, the largest optical instrument in Ukraine, and the Crimean Astrophysical Observatory (CAO), one of the country's [four World's Astronomy Monuments](#), known for its prolific contributions to comet and [minor planet detections](#). The CAO served as an observation base for many Ukrainian astronomers, including researchers of variable stars, small bodies of the Solar System, galaxies with active nuclei, and a platform for student training activities. A few other powerful Ukrainian telescopes, such as the GT-48 gamma-ray telescope, radiotelescopes P-400P (32m) and RT-70 (70m), are also located in Crimea. These facilities are now inaccessible to Ukrainian scientific institutions, and their losses were an early major disruption to the astronomical research programs. Ukraine also lost access to the Terskol Observatories in Russia's northern Caucasus. Part of the International Centre for Astronomical, Medical, and Ecological Research (ICAMER), the telescopes were jointly operated by the Ukrainian and Russian Academies of Sciences.

Among Ukraine's most important research institutions is the Main Astronomical Observatory (MAO) of the National Academy of Sciences of Ukraine (NASU) in Kyiv. In June 2025, its central building was damaged by shock waves from blasts nearby, as was the structure housing the ACU-5 horizontal solar telescope. Another crucial observation and student academic training station in Mayaky, belonging to the Astronomical Observatory of Odesa I.I. Mechnikov National University, is not working because it is located in a high-risk area. While equipment in these sites remains intact, other observatories — especially in Kharkiv — were heavily impacted.

Up until January 2025, at least six Ukrainian regions registered substantial damage to educational facilities, and Kharkiv was the most severely hit. Since the beginning of the war, astronomical observatories in the area have been subject to heavy damage, widespread destruction, and looting. Located 70 km to the southeast of the city of Kharkiv, the Chuhuiv Observational Station of the V. N. Karazin Kharkiv National University was occupied by the Russian army and pro-Russian military groups from Luhansk and Donetsk. Over six months, critical equipment—computers, data storage systems, control units, power supplies, CCD cameras, and electric drives,—was looted or destroyed. Shrapnel also damaged the AZT-8 (70 cm) and 40-cm survey telescope domes. At the meeting in Leiden, Vadym Kaydash, director of the Astronomy Research Institute at the Kharkiv National University, showed images of destroyed instruments and pictures of shells and missiles landed near these facilities (see Fig. 1 for a multiple launch rocket system combat unit found near Kharkiv University)[3]. Demining only began in July 2025.

**Fig. 1: Evidence of military action near academic facilities.**

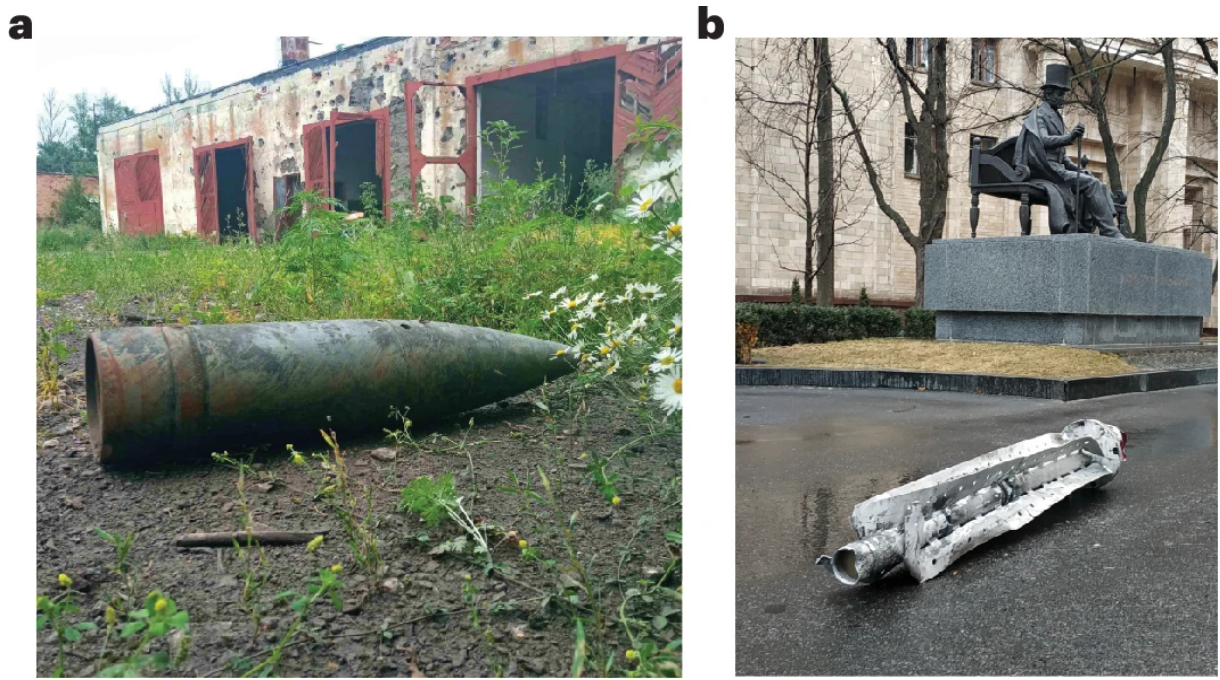

**a**, A 120 mm tank shell that landed near the Braude Radio Astronomical Observatory. **b**, Part of a multiple launch rocket system unit in front of Kharkiv University and broken windows due to the shockwave in the background. Credit: Anatolii Pryvalov.

The Braude Radio Astronomical Observatory, home to the world's largest low-frequency radiotelescope UTR-2, located in the Kharkiv region, was used by the occupiers as a military base for six months in 2022. As a result, 16 of the 17 buildings on the observatory's territory were substantially damaged, and one house was completely destroyed (see Fig. 2). Servers and other computer equipment were looted, the phasing system and part of the cables were destroyed or stolen. Fortunately, only a few dozen of the 2,040 antennas of the UTR-2 radio telescope were damaged (see Fig. 3 for an example of a broken antenna). The site is also home to the Giant Ukrainian Radio Telescope (GURT), with 275 antennas, of which no more than 50 were found to be completely intact (see figure 2 for an example of a broken antenna). After extensive repairs, Ukrainian radioastronomers reported that GURT's current status is now back to operational[4].

**Fig. 2: Damage to the buildings at Braude Radio Astronomical Observatory.**

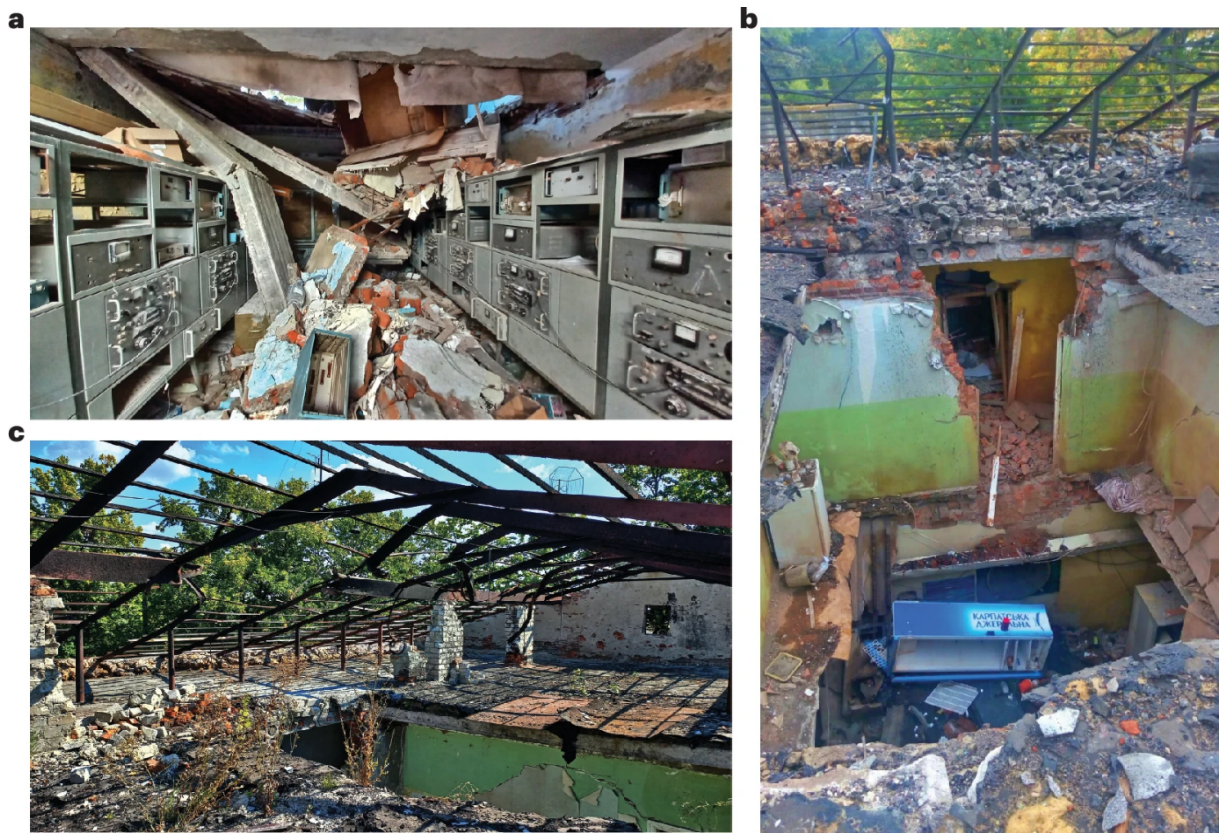

**a**, Destroyed equipment room of the Braude Radio Astronomical Observatory. **b**, Top-down view of a destroyed building at the Braude Radio Astronomical Observatory. **c**, Roof view of destroyed building at the Braude Radio Astronomical Observatory. Credit: Mikhail Sidorchuk.

**Fig. 3: A broken radio antenna at the Braude Radio Astronomical Observatory.**

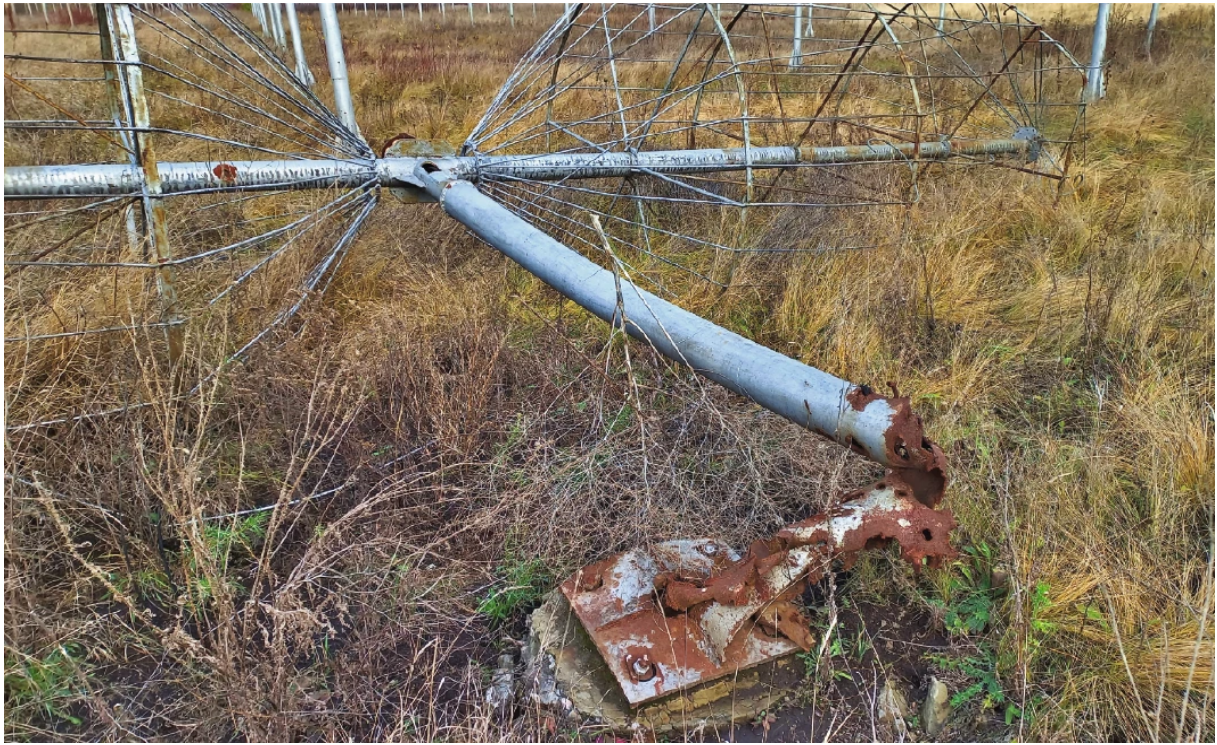

One of the broken antennas of the UTR-2 at Braude Radio Astronomical Observatory. Credit: Mikhail Sidorchuk.

Being forced to live abroad is an outcome far too many Ukrainian astronomers have experienced since the war started, and is part of a broader social dimension of the damages to science caused by the Russian invasion: researchers subject to exile or internal displacement or work under rocket-drone military attacks in Kyiv, Kharkiv, Odesa, Dnipro and other cities, where the majority of Ukrainian astronomers continue research.

**Brain-drain and internal displacement**

The war has substantially disrupted the Ukrainian astronomy workforce. According to Igor Lukianyk, deputy director for research at the Astronomical Observatory of the Shevchenko National University of Kyiv, the institution experienced sharp declines in the number of researchers: the total staff today is 56% smaller compared to 2022 (Fig. 4)[8]. Lukianyk reported that three staff members are currently serving in the Ukrainian armed forces[8]. Similar cases are noted in institutions from Odesa, Lviv and other centers[6, 7]. At the Research Institute of Astronomy of Kharkiv National University, 75% of researchers and teachers were displaced during the occupation in 2022[3]. According to Kaydash, as of 2025, up to 20% of researchers remain abroad, and half of the students have been relocated[3]. In January 2024, scientists from the Ukrainian wartime diaspora were scattered across 53 countries, with more than half of 1,509 relocated researchers being absorbed by institutions in Germany, Poland, the UK and France[1].

**Fig. 4: The decline in staff at Ukrainian astronomical institutions.**

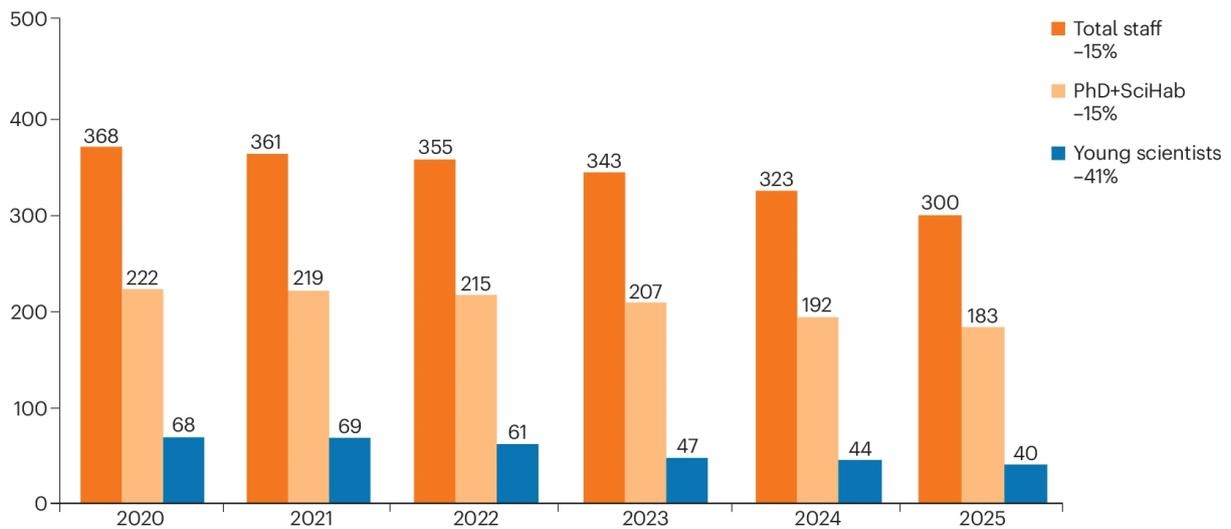

Decline in research personnel at Ukrainian astronomy institutions between 2020–2025, according to data gathered by the Ukrainian Astronomical Association. 'Young scientists' refers to early-career researchers under 35 without a scientific degree or with a PhD degree and under 40 in the case of 'SciHab'. 'PhD+SciHab' includes staff with a PhD degree or scientific Habilitation degree. 'Total staff' encompasses both groups as well as other research-related personnel, including engineers or employees without scientific degrees who conduct research or hold senior research positions. The sharpest decline is observed among 'Young scientists' (–41%).

Dozens of researchers have received temporary placements abroad through the Marie Skłodowska-Curie Actions (MSCA) funded MSCA4Ukraine project, PAUSE (the French hosting program for scientists and artists in exile), and national grants from several European countries[8]. While these fellowships have helped maintain scientific careers, there is a growing risk of permanent brain-drain unless reintegration is actively supported. The Ukrainian Astronomical Association reported that only 40 of its 300 current members are early-career scientists[8]. Reversing this trend will require a coordinated effort in supporting training, mobility, and international collaboration. The EURIZON (European scientific and technical collaboration in the field of research infrastructures) initiative has provided support to Ukrainian scientists, enabling them to continue research despite the war.

Student enrollment has declined nationwide. Kyiv, Kharkiv, Odesa and Lviv universities report fewer applicants and graduates in astronomy, citing waning public interest and interrupted schooling[3, 5]. Frequent air-raid alarms and infrastructure damage have forced classes online or into shelters, limiting access to observatories and laboratory-based training. In parallel, large-scale emigration—particularly of families with children—has led to a national decline in school enrolment and a disruption in the pipeline of future students entering physics and astronomy.

The Ukrainian Astronomical Association (UAA) has played a key role in preserving institutional continuity, integrating displaced researchers, and supporting early-career scientists through conference funding, awards, and textbook publication in Ukrainian. Mergers of observatories and realignment under the NASU have also helped safeguard critical personnel and research infrastructure.

**Looking ahead**

While the world scrambles to put an end to the war with still no foreseeable diplomatic solution, the plan to redevelop Ukrainian astronomy with a deeper integration into European and the wider international community started being discussed during the meeting *"Recovery Plan for Ukrainian Astronomy"*. Organized by Leiden University and the European Regional Office of Astronomy for Development of the International Astronomical Union (IAU-EROAD), the event opened the dialogue between Ukrainian, Dutch, and European astronomical communities to develop a strategic recovery framework to support Ukrainian astronomy both during the war and in the post-war period. Future plans include repairing damaged facilities, reconstructing destroyed research infrastructure, implementing educational exchange programs, and strengthening collaborative partnerships between Ukrainian astronomers and their European and international colleagues.

Details of the initial plan discussed in Leiden are only going to be published in November, but provisional results from workgroups during the meeting anticipate that a recovery of Ukrainian astronomy should not aim exclusively at rebuilding destroyed structures and replacing lost equipment, although this should not be overlooked, either. Most importantly, it means looking ahead: fostering innovation and economic development through astronomy was one of the most detailed guidelines drafted during the event, providing a 10-year roadmap to integrate astronomy and the industrial sector. It includes plans for launching industry career training programs for students, to create innovation hubs and incubators, and deploy initial technology spin-offs in astronomy.

On the human side, more investments in partnerships between European institutions and the Ukrainian astronomy community were deemed paramount for the recovery process, building upon success stories like the current participation of Ukraine in the ACME (Astrophysics Centre for Multi-messenger studies in Europe) and EURIZON projects. Workgroups also proposed to enhance networking between Ukrainian and European institutions with possible bilateral cooperation and fellowship programs and mentorship programs involving Ukrainian students working together with foreign mentors and vice-versa. Additionally, a school for young researchers is being organized in Ukraine during wartime, offering training and mentorship to students, with support from small businesses committed to sustaining scientific education under difficult conditions.

In late 2023, a Ukrainian radioastronomer commented on the uncertain future of the damaged UTR-2 observatory: "There's no point in restoring it as it was. We want to modernize it, but only after the war."[11] Maybe that's now somewhat applicable to the whole field of astronomy in Ukraine: the still murky and dangerous present poses limitations to recovery initiatives, but plans for a modern and more internationally integrated Ukrainian astronomical community should be ready once post-war civilian reconstruction efforts get on track. In this context, the most important aspect is that Ukrainian astronomers kept playing their part. Paraphrasing Kaydash, Ukrainian astronomy may have been temporarily blinded[3], but its community continued moving forward even in face of many losses and disruptions, in the most challenging circumstances.

**Acknowledgements**

We'd like to thank the following speakers at the "Recovery Plan for Ukrainian Astronomy" meeting: Michiel Rodenhuis (Executive director, Netherlands Research School for Astronomy, NOVA), Michael Wise (Director general, Space Research Organisation Netherlands, SRON), Violette Impellizzeri (Head of Astronomy & Operations, Netherlands Institute for Radio Astronomy, ASTRON), Michiel van

Haarlem (Executive director, LOFAR ERIC), Giuseppe Cimo (Head of Space Science and Innovative Applications at JIV-ERIC), Sara Lucatello (President, European Astronomical Society, EAS), Saskia Matheussen (Teamleader of the Dutch Research Council, NWO, Deputy Chair of ASTRONET), Pavlo Plotko (Scientific coordinator, German Center for Astrophysics, DZA, in formation c/o Deutsches Elektronen-Synchrotron, DESY), Barbara Dammers-Szenasi (Policy officer, Dutch Research Council, NWO).

**Competing Interests**

The authors declare no competing interests.